\documentclass[runningheads]{chaos}

\usepackage{graphicx} % standard LaTeX graphics tool
\usepackage{latexsym}
\usepackage{amssymb}
\usepackage{bm}
\newcommand{\f}{\begin{equation}}
\newcommand{\ff}{\end{equation}}

\begin{document}
%
% \title* lets you specify the title of your manuscript.
% Use \protect\newline to force a line break in your title.
\title*{Gauge Field Turbulence as a Cause of Inflation in Chern-Simons Modified Gravity}
%
% \toctitle specifies the title as will be printed in the table of 
% contents.
% Use \protect\newline to force a line break in your title.
\toctitle{Gauge Field Turbulence as a Cause of Inflation in Chern-Simons Modified Gravity}
%
% \titlerunning defines the title in the running head. Abbreviate
% your title, if the full title is too long to fit in the running 
% head.
\titlerunning{Gauge Field Turbulence as a Cause of Inflation in Chern-Simons Modified Gravity}
%  
% \authors specifies the authors. Please use initials. Authors are 
% seperated by the \and command. Use the \inst{1} and \inst{2} commands 
% to define the reference mark to your affiliation if needed.
\author{
  David Garrison\inst{1}
}
%
% The following command allow each of the authors to appear 
% in the author index.
% \index{Author, A.}
\index{Garrison, D.}

%
% \authorrunning specifies the author name(s) in the running head.
% If there are more than two authors, please abbreviate author list
% (e.g., Skiadas  et al.) for running head.
\authorrunning{Garrison}
%
% The \institute command lets you specify  your affiliation and
% your address. Seperate two or more different affiliations by the
% \and command. 
\institute{
  University of Houston-Clear Lake, 2700 Bay Area Blvd., Houston, TX, USA\\
  (E-mail: {\tt garrison@uhcl.edu})
}

% Typeset the title
\maketitle             

\begin{abstract}
In this paper, we study the dynamics of the Chern-Simons Inflation Model proposed by Alexander, Marciano and Spergel.  According to this model, inflation begins when a fermion current interacts with a turbulent gauge field in a space larger than some critical size.  This mechanism appears to work by driving energy from the initial random spectrum into a narrow band of frequencies, similar to the inverse energy cascade seen in MHD turbulence.  In this work we focus on the dynamics of the interaction using phase diagrams and a thorough analysis of the evolution equations.  We show that in this model inflation is caused by an over-damped harmonic oscillator driving waves in the gauge field at their resonance frequency.\\
\end{abstract}

\section{Introduction}

According to accepted cosmological theory, the early universe went through a period of inflation where it's size increased by at least 60 e-folds in a small fraction of a second \cite{guth}.  There is a wealth of observational data providing evidence that inflation occurred \cite{Ade:2014xna}.  In addition, Inflation is needed to satisfy several fundamental problems in cosmology such as the flatness and horizon problems.  Unfortunately, most theories of inflation involve the existence of a scalar field and are difficult to distinguish by observation or experiment.  Also, there are still many unanswered questions about where the scalar field came from or why it disappeared after inflation ended.  Recently, Alexander et al. \cite{alexander1} suggested a new theory of cosmic inflation based on Chern-Simons modified gravity \cite{alexander2}  Unlike scalar field theories of inflation, the theory proposed by Alexander et al. utilizes the interaction between a gauge field and fermion current to drive inflation and does not depend on the existence of a scalar field.  This is not the only theory of inflation derived from a vector field interaction \cite{ford} but it is unique in that it involves elements that are known to exist in practice and not just in theory.  

The Chern-Simons Inflation theory works by suggesting that the energy density from the interaction between the gauge field and fermion current behaves like vacuum energy.  This is possible in Chern-Simons modified gravity.  The gauge field starts with a random white noise spectrum but then the energy is transported to a few low frequency modes.  In the early version of the paper by Alexander et al, the spatial parts of the gauge field and fermion current where used to derive the energy density.  They later changed that and based the energy density on the temporal parts of the gauge field and fermion current.  This author believes that the motivation for this change may have been the belief that the spatial part of the fermion current dropped to zero too quickly to be effective in driving inflation.  We find this to not be the case.  We also find that the temporal part of the gauge field and fermion current may not be sufficient to drive a 60 e-fold increase in scale factor.

Our code utilizes the Adler-Bell-Jackiw (ABJ) chiral anomaly \cite{adler,bell} to model the decrease in fermion current associated with changes in the scale factor and gauge field.  This is a small quantum mechanical violation of the conservation of axial-vector current.  This violation occurs due to tunneling of fermions from one vacuum to another and is partially responsible for the gentle ending of the inflation event.  It is the means by which the gauge field converts to leptons during inflation resulting in lepto-genesis.  As the current decreases during inflation, the negative pressure driving inflation should decrease as well unless the decrease in current is offset by an increase in the gauge field.

The overall goal of the study presented here is to understand the dynamics of the system and strengthen our physical interpretation of the theory presented here.  The author's previous paper on the Numerical Simulation of Chern-Simons Inflation \cite{garrison} served to prove the feasibility of the theory.  Additional work is being planned to more thoroughly study the version of the theory involving the temporal part of the gauge field and fermion current both alone and in conjunction with the spatial part.  In the following sections, we will describe in more detail what we believe is the most promising model of Chern-Simons Inflation as well as the results of computer simulations of this model.  In the final section, we will discuss these results and how they may be used in future research.

\section{Model and Simulations}

The code utilized in these simulations is based on the Cactus framework \cite{goodale} used for Numerical Relativity research.  While Cactus is an extremely sophisticated code containing millions of lines of code, all the physics is contained in code written by the author.  This code has been thoroughly tested and the results are self-consistent and reliable.

The inflation model developed by Alexander, Marciano and Spergel utilizes a gauge field which interacts with fermions in the early universe to produce an effective scalar field that generates inflation~\cite{alexander1}.  See the recent article by Garrison and Underwood for a complete description of how the numerical equations for this model are derived \cite{garrison}.

For the numerical calculation, we use natural units but later evaluate the data in terms of SI units so that the results can be easily compared to the established values. In order to use this model in our code, we separated the equations of motion for the gauge field, ABJ chiral anomaly, Chern-Simons term and the Friedman equations into a system of first order in time differential equations.  
\begin{eqnarray}
\frac {d \vec A} {dt} = && \frac {\vec Z} {a} \\
\frac {d \vec Z} {dt} = && \vec J a^3 + \nabla^2 \vec A / a - a^2  \frac {\dot{\theta}} {M_{*}}  \vec B \\
\frac {d J^0} {dt} =  && \frac{\vec E \cdot \vec B}{4{\pi ^2}{a^2}} - \vec \nabla \cdot \vec J - 2H J^0 \\
\frac{d D}{dt} = && \frac{\vec E \cdot \vec B}{4a^3 M_*^2} - 3 H D - 2 \frac{m^2} {M_*} \theta \\
\frac{d\theta}{dt} = && D M_* \\
\frac {d a} {dt} = && a H \\
\frac {d H} {dt} = && \frac {8 \pi} {3} \bar{\rho} / a - H^2
\end{eqnarray}
Here the gauge field is represented with $\vec A$, the current is $\vec J$, a is the scale factor and H is the Hubble parameter.  Current is assumed to depend simply on the charge density according to the equation $\vec J = J^0 \vec v$.  $\vec E$ represents the hyper charged electric field, $\vec E \equiv \vec {\dot{A}}$.  $\vec B$ is the hyper charged magnetic field, $\vec B \equiv \nabla \times \vec A$ term.  $M_{*}$ is the mass scale identified with the UV cut-off scale of the effective field theory and $\theta$ is responsible for CP violation.  $m$ is on the order of the GUT energy scale.  Finally, The average energy density is calculated as $\bar{\rho} = \frac{1}{N} \sum_{k=1}^N \frac {E_{k}^2 + B_{k}^2} {2 a^4} + |\vec A_k \cdot \vec J_{k} / a|$.  Here N represents the total number of grid points in the computational domain.  The scale factor and Hubble parameter therefore depend on the average energy density and not the local field dynamics. 

The initial gauge field was composed of a random (white noise) spectrum.  In order to generate the initial gauge field, we used a random number generator to create a random spectrum with amplitude up to the calculated maximum amplitude, $|\vec A|$, in each direction.  The magnitude of the gauge field was then held equal to the initial amplitude $|\vec A|$.  The initial values for the variables used in this study are given below.
\begin{eqnarray}
| \vec A | = && 1.0 \times 10^{-5}M_{P} \\
J^0 = && 10^{-10} M_{P}^3 \\
\vec{v} = && 10^{-10} (\hat{x} + \hat{y} + \hat{z}) \\
\frac {\dot{\theta}} {M_{*}} = && 2.18 \times 10^{-5} M_{P}  \\
a = && 1.0 \\
H = && \sqrt{\frac{8 \pi} {3} | \vec A \cdot \vec J |} \\
\vec Z = && H a \times random~number(-1,1) \\
m = && 4.15 \times 10^{-6}M_{P} \\
M_* = && 4.15 \times 10^{-6}M_{P} 
\end{eqnarray}
The code was then run on the University of Houston$\textquoteright$s Maxwell cluster using a variety of time-steps, grid sizes and resolutions in order to obtain consistent results.   A fourth order finite differencing scheme was used to test convergence for high and low resolution simulations.  Because the initial units were entered as Planck units, we assumed that the physical grid (horizon) size corresponded to Planck lengths and the timing output could be interpreted as Planck time. 

\section{Results}

The previous article by Garrison and Underwood \cite{garrison} focused on demonstrating the feasibility of the model and verifying that the apparent inverse energy cascade occurred as predicted.  Previous data have shown that this is an interesting chaotic system which is highly dependent on initial conditions but numerically stable for a large range of initial data.  As in the previous paper, Figure 1 shows the life-cycle of the evolution as our virtual universe experiences inflation.  This is demonstrated by the scale factor and Hubble parameter.  

\begin{figure}
\includegraphics[height=2.5in,width=4.75in,angle=0]{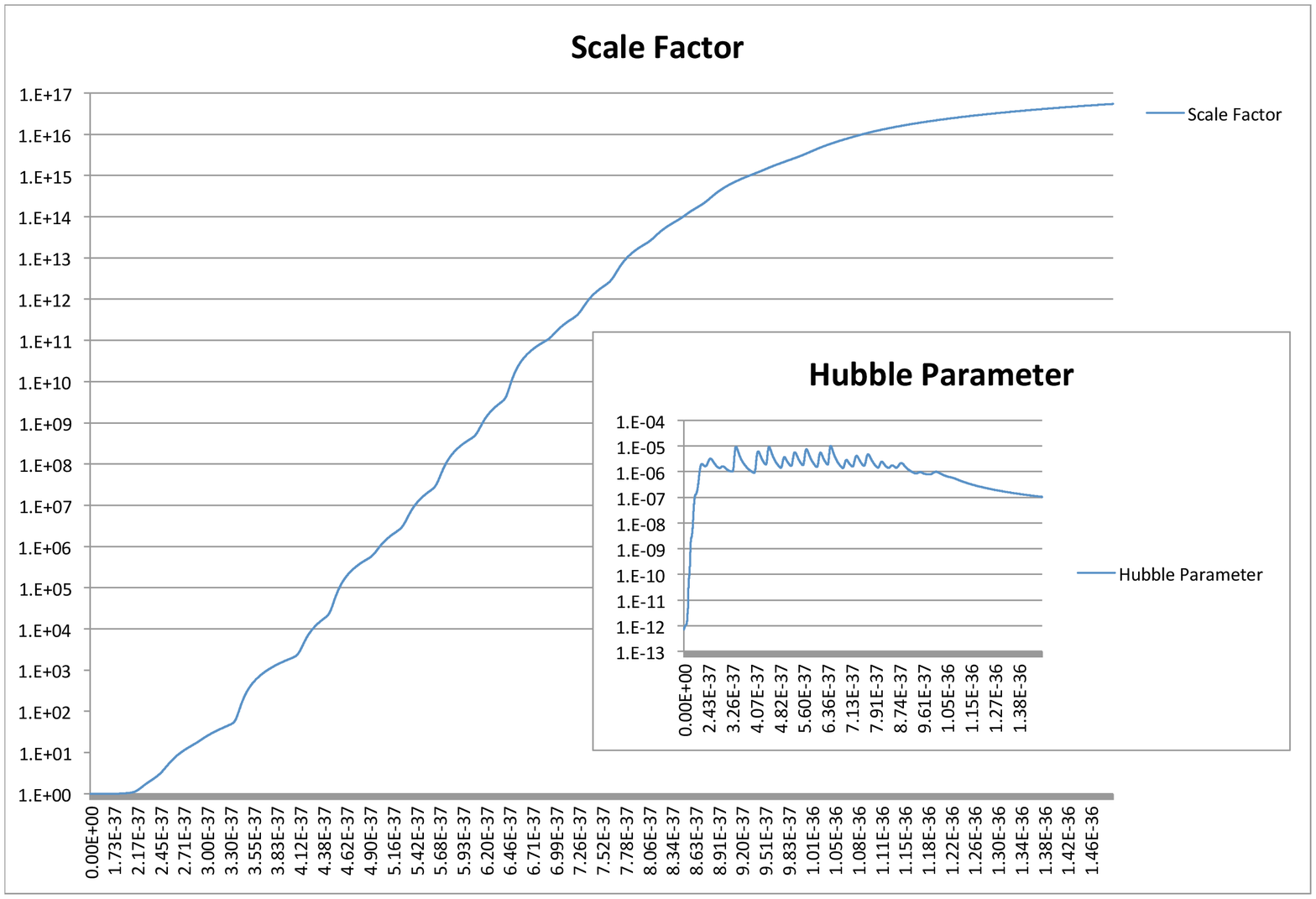}
\caption{The scale factor and Hubble Parameter for the inflationary period (log scale).\label{fig:scale}} 
\end{figure}

Figure 2 shows how the gauge field increases and charge density decreases with time.  The net result of this is that the dot product of the gauge field and charge density yields a nearly constant energy density (and therefore Hubble parameter) until inflation ends.  The gauge field evolution equation is essentially an inhomogenous  wave equation driven by the $\frac {\dot{\theta}} {M_{*}}  \vec B$ term and the $\vec J a^3$ term.  Understanding how inflation occurs is directly connected to the dynamics of these two terms.  Charge density falls off as roughly $a^{-2}$ making the second term simply increase proportional to the scale factor and therefore lack the dynamics needed to significantly effect the gauge field evolution.  The first term however is much more dynamic and could explain why inflation begins and ends.  Also, given our initial conditions, the first term is $\sim10^{-5} \vec B$ while the second term is $\sim10^{-20}$ so the first term should normally dominate since $\vec B$ starts around $10^{-9}$ and increases as quickly as the gauge field.

\begin{figure}
\includegraphics[height=2.5in,width=4.75in,angle=0]{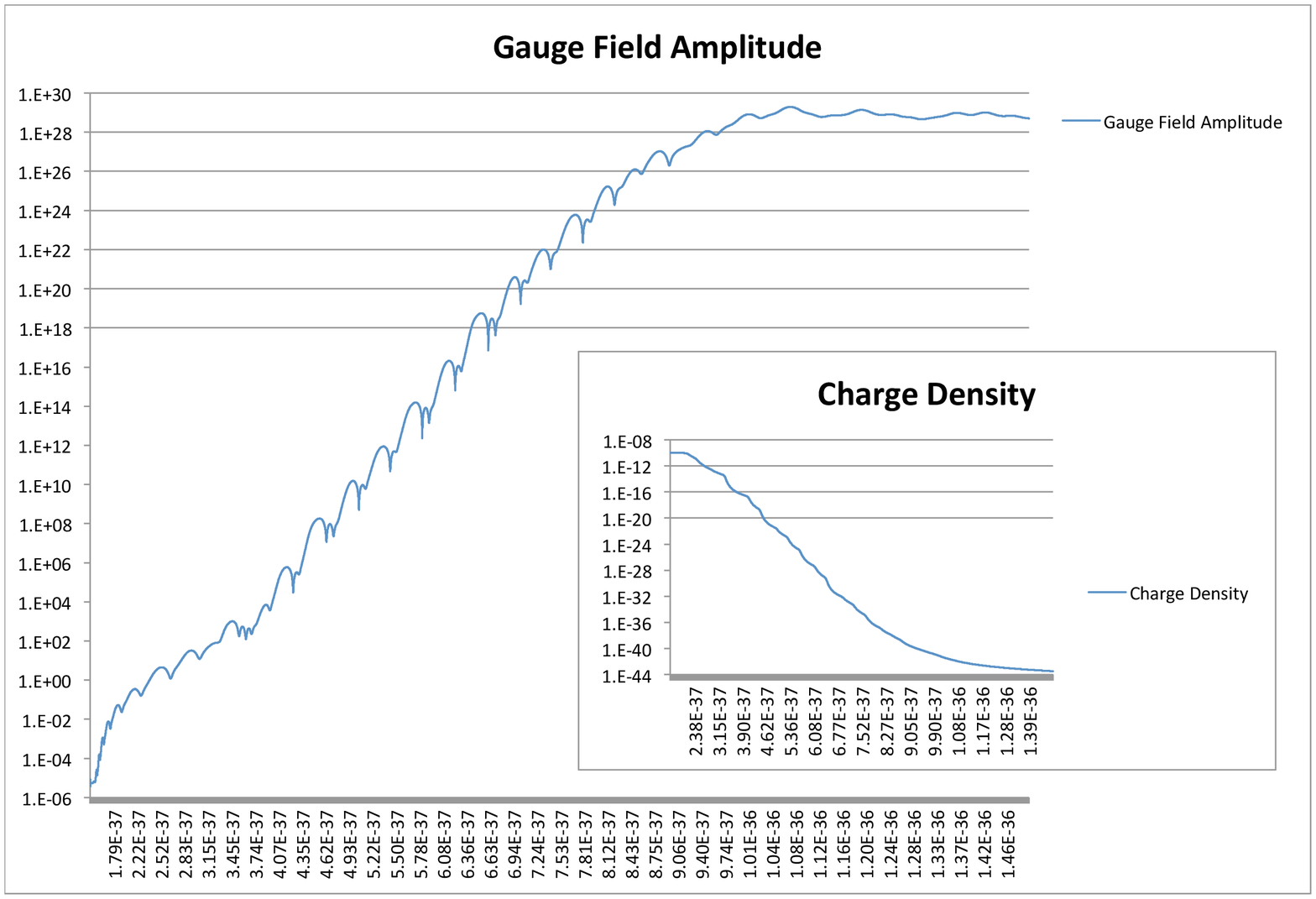}
\caption{The Gauge Field Amplitude and Charge Density (log scale).\label{fig:gauge}} 
\end{figure}

Without the $\theta$ term Chern Simons modified gravity reduces to ordinary General Relativity and the effective vacuum energy disappears.  Much of the gauge field dynamics is therefore the result of the $\theta$ term and it's time derivatives.  The phase diagram in Figure 3 shows that this term acts like a dampened driven harmonic oscillator.  The frequency of this system is $\omega = \sqrt{2}  m$ the dampening term is $\gamma = \frac{3}{2} H$ and the driving term is F = $\frac{\vec E \cdot \vec B}{4a^3 M_*}$.  Given our initial conditions, $\omega \approx 10^{-6}$, $\gamma \approx 10^{-12} \rightarrow 10^{-5}$ and F is insignificant because $\vec E \cdot \vec B$ is unmeasurably small.  This is therefore an under-damped harmonic oscillator that transitions into an over-damped harmonic oscillator as the Hubble parameter increases.  The $\theta$ term vanishes quickly after $\gamma$ exceeds $\omega$ and the gauge field's rate of growth slows while current continues to decrease at a constant rate resulting in a decreasing energy density and an end to inflation.  Maintaining the Chern Simons term, $\frac {\dot{\theta}} {M_{*}}$, for as long as possible appears to be essential to the inflation process. 

\begin{figure}
\includegraphics[height=2.5in,width=4.75in,angle=0]{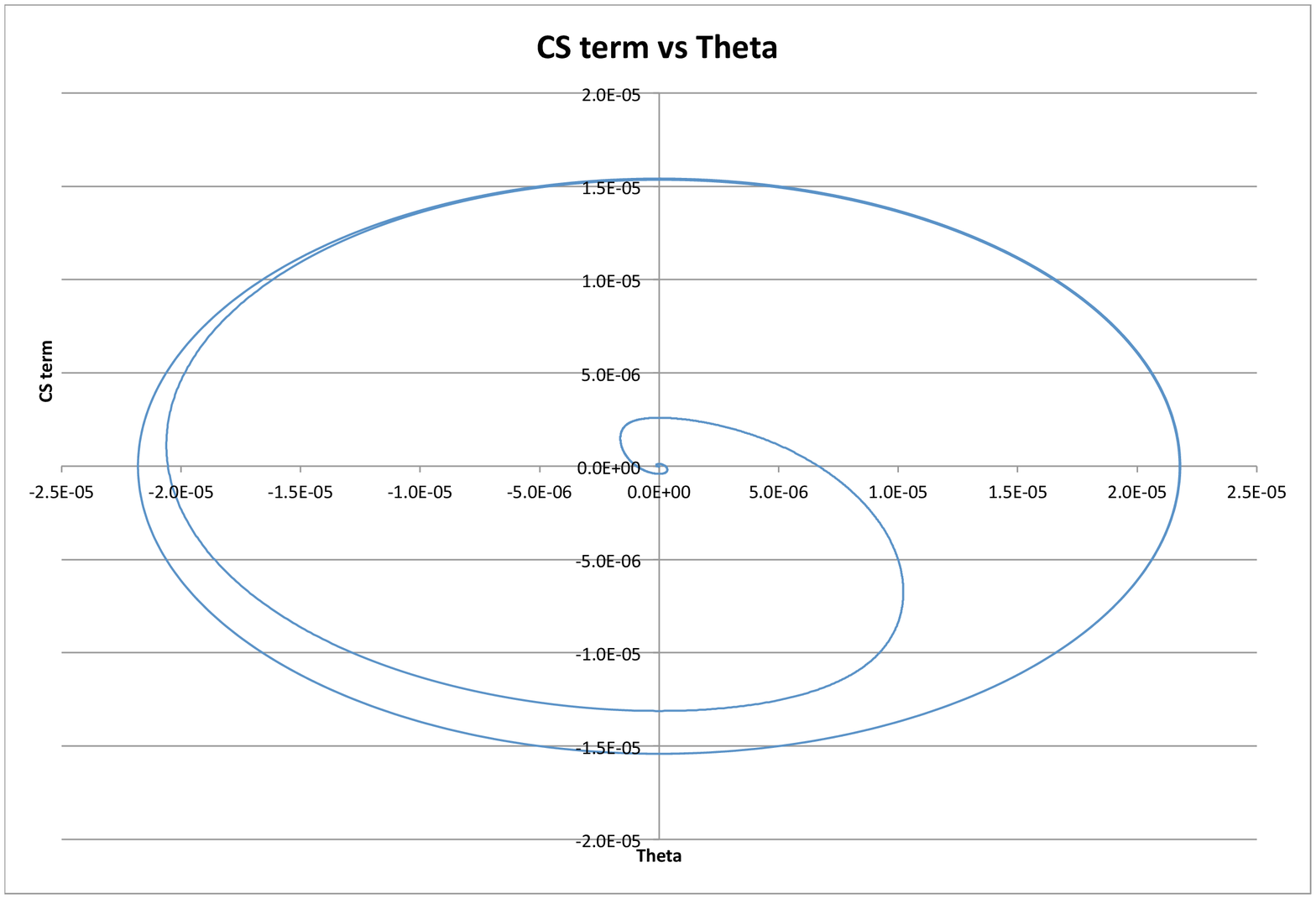}
\caption{Phase diagram of the $\theta$ and $\frac {\dot{\theta}} {M_{*}}$ terms.\label{fig:theta}} 
\end{figure}

In Figure 4, we see the spectrum of the gauge field as a function of time.  Notice that the initial gauge field starts off with an evenly distributed spectrum and then sometime later the spectrum peaks at low frequencies to resemble an inverse energy cascade.  Later the peak of the power spectrum moves to higher frequencies as the Chern Simons term decays.  The peak follows the changing frequency of the Chern Simons term to maintain resonance until it decays to zero and inflation ends.

\begin{figure}
\includegraphics[height=3.5in,width=4.75in,angle=0]{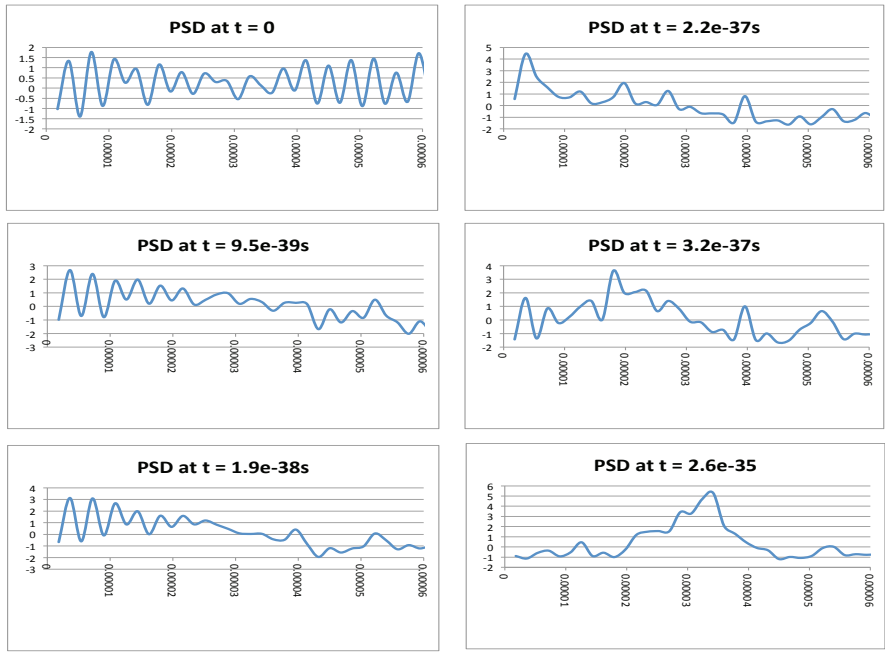}
\caption{Power Spectral Density of the gauge field at various times.  The vertical axis is the log of the deviation from the mean.  The horizontal axis is frequency in Plank units. \label{fig:PSD}} 
\end{figure}

\section{Conclusions}

An important result of this study is that we now know why inflation only appears to occur when the computational grid is sufficiently large.  Our analysis of the $\theta$ term's dynamics show that it's natural oscillatory frequency is on the order of $10^{-6}$.  This corresponds to a grid size of about $10^6$ units.  If the computational grid is smaller than this minimum, the gauge field cannot come into resonance with the driver $\theta$ and explode in amplitude.  

Information from this study may also be useful in better determining what initial conditions led to inflation in our universe.  By varying the initial conditions, the forcing term, dampening term and frequency of $\theta$ may be altered to extend our simulated inflation and better conform to observation.

\end{document}